\documentclass[10pt]{iopart}
\usepackage{graphicx}
\begin{document}
\title{A novel concept for increasing the peak sensitivity of LIGO by detuning the arm cavities}
\author{S. Hild$^1$ and A. Freise$^2$}
\address{$^1$ Max-Planck-Institut f\"ur Gravitationsphysik
(Albert-Einstein-Institut) and Leibniz Universit\"at Hannover,
Callinstr. 38, D--30167 Hannover, Germany.}
\address{$^2$ School of Physics and Astronomy, University of
Birmingham,   Edgbaston, Birmingham, B15 2TT, UK}
\ead{stefan.hild@aei.mpg.de}

\begin{abstract}
We introduce a concept that uses detuned arm cavities
 to increase the shot noise limited sensitivity of  LIGO
without increasing the light power inside the arm cavities.
Numerical simulations show an increased sensitivity between 125
and 400\,Hz, with a maximal improvement of about 80\,\% around
225\,Hz, while the sensitivity above 400\,Hz is decreased.
Furthermore our concept is found to give a sensitivity similar to
that of a conventional RSE configuration with a Signal-Recycling
mirror of moderate reflectivity. In the near future detuned arm
cavities might be a beneficial alternative to RSE, due the
potentially less hardware intensive implementation of the proposed
concept.

\end{abstract}


\section{Introduction}

The first generation of large-scale laser-interferometric
gravitational wave detectors \cite{geo, virgo, ligo, tama} is now
in operation and collects data of impressive sensitivity and
bandwidth. The optical configurations of these kilometer long
gravitational wave observatories are based on a Michelson
interferometer. Moreover, the standard configuration, implemented
in the three LIGO interferometers as well as in Virgo  and TAMA300,
employs cavities in the arms of the interferometer and Power-Recycling
to increase the storage time of the light inside the
interferometer. In order to get an optimal power increase inside the arm
cavities, these are kept to be resonant for the carrier light.
However, as we show in this article, detuning these arm cavities
by a few hundred Hz has the advantage of a larger signal gain in a
certain frequency band and might therefore  be favorable.

In section \ref{sec:principle} we give a brief and intuitive
description of the principle of an optical configuration employing
detuned arm cavities. Using a idealized interferometer as an
example we show in section \ref{sec:simulation} that with detuned
arm cavities the shot noise limited sensitivity can, in a certain
frequency range, be increased using the same optical power
circulating in the arm cavities as in the case of tuned arm
cavities. In section \ref{sec:rse} we compare the interferometer
response with detuned arm cavities to a configuration using in
addition the advanced technology of resonant sideband extraction
(RSE). The potential benefit of using detuned arm cavities for
LIGO is evaluated by simulations described in section
\ref{sec:real_LIGO}. Finally we give a summary and an outlook in
section \ref{sec:summary}.

\section{The principle of detuned arm cavities}
\label{sec:principle}

\begin{figure}[Htb]
\centering
\includegraphics[width=1\textwidth]{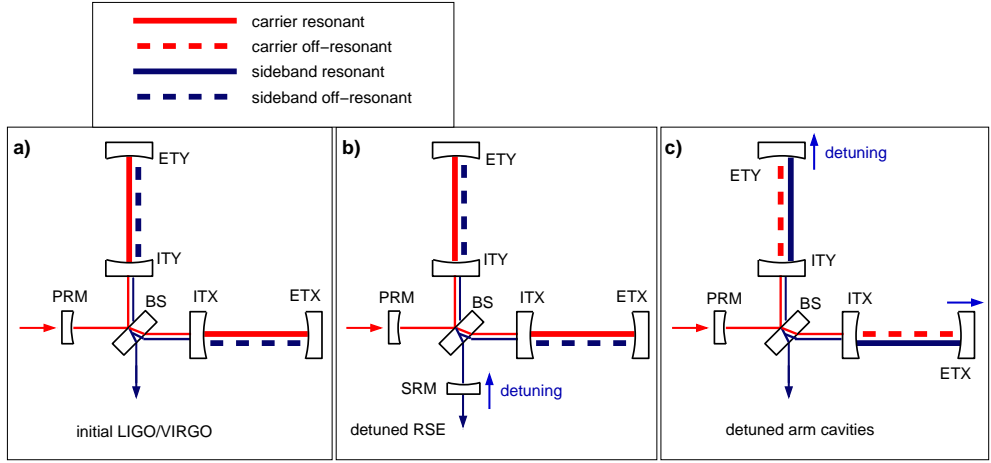}
\caption{Simplified schematics of the three optical configurations
compared in this article. a) Initial LIGO configuration with arm
cavities kept resonant for the carrier light. b) Initial LIGO
configuration with additional detuned resonant sideband extraction. c)
Initial LIGO configuration with arm cavities detuned to be
resonant for the gravitational wave signal sidebands
(abbreviations are explained in the text.)} \label{layouts}
\end{figure}

Figure \ref{layouts} a) shows a simplified schematic of the
initial LIGO optical layout. The red and the dark blue lines indicate
in which part of the interferometer carrier light (red) and
gravitational wave signal sidebands (dark blue) are present. The
carrier light enters the interferometer through the Power
Recycling mirror (PRM) and is split by the beam splitter (BS) in
equal shares into the X and Y arm. The arms consist of an arm
cavity each, formed by two mirrors separated by 4\,km (ITX and
ETX, ITY and ETY). These arm cavities are chosen to be resonant
for the carrier light resulting in a larger power enhancement
inside. The small Michelson interferometer formed by BS, ITY and
ITX is kept on a dark fringe, thus no carrier light is leaving the
interferometer towards the output port, but all light is going
back to the input port so that it becomes further resonantly enhanced by
PRM. The presence of a gravitational wave produces phase
modulation sidebands around the carrier light. Because the signal
sidebands in the two perpendicular arms have opposite phase they
can interfere constructively at the BS to leave the interferometer
at the output port. With the arm cavities being set on resonance
for the carrier light, the signal sidebands of interest
($f_{sig}>100$\,Hz) experience less enhancement than the carrier light.

A new concept that turns this principle around is to use detuned
arm cavities\footnote{Similar concepts were mentioned in
\cite{malik} and \cite{rehbein}; their publications, however,
focused on the properties of the opto-mechanical resonance in a
detuned arm cavity, while in this paper we investigate the
sensitivity enhancement from utilising the optical resonance.} as
shown in Figure \ref{layouts} c). In contrast to the conventional
initial LIGO scheme the length of the arm cavities was chosen to
be resonant for a specific single-sided signal sideband frequency,
i.e. the cavities are detuned from the carrier frequency. This
detuning can easiest be realized by shifting ETY and ETX in common
mode either to shorten or to lengthen the arm cavities,
corresponding to making them resonant for the upper and the lower
signal sideband, respectively. The operating points of the other
main mirrors (ITX, ITY, and BS) do not need to be changed. The
obvious drawback of this scheme, in the following referred to as
\emph{detuned arm cavities}, is the lower optical power inside the
arms, compared to arm cavities resonant for the carrier light.
However, the achievable circulating power in current detectors is
not limited by the available laser power nor the finesse of the
arm cavities. Instead technical problems, like, for example,
thermal lensing or parametric instabilities, become limiting
factors when the circulating light power is increased \cite{ligo}.
Therefore we belief it useful to compare the concept of detuned
arm cavities to resonant arm cavities with identical stored
optical power.

%

\section{Simulated shot noise limited sensitivity for an ideal
Michelson interferometer employing Power-Recycling  and detuned
arm cavities} \label{sec:simulation} 
\begin{figure}[Htb]
\centering
\includegraphics[width=1\textwidth]{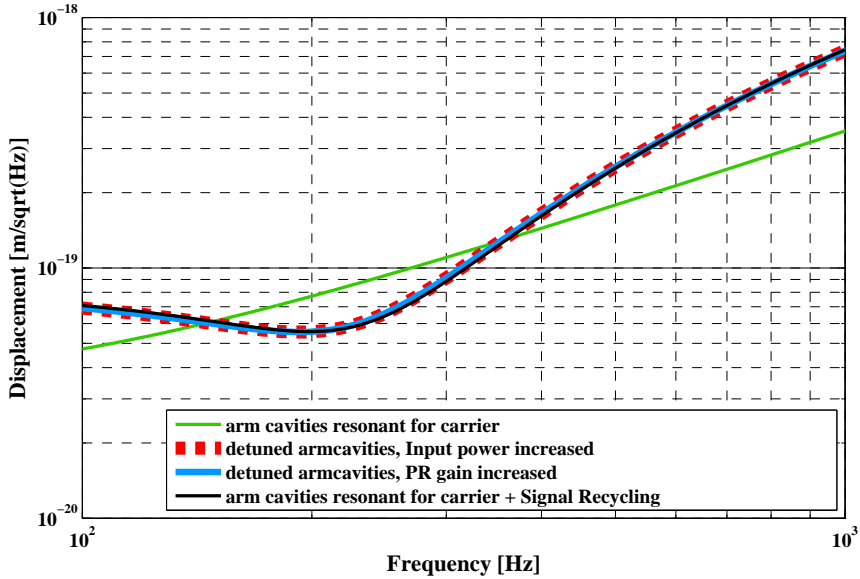}
\caption{Shot noise limited displacement sensitivity of an ideal
Michelson interferometer with Power-Recycling with DC-readout in
comparison to two configurations with detuned arms and a
configuration using resonant sideband extraction (RSE). Using
detuned arm cavities increases the sensitivity in a frequency
range from 150 to 350\,Hz. Detuned arm cavities give a sensitivity
identical to RSE.} \label{Initial}
\end{figure}

In order to evaluate the
benefit from detuning the arm cavities and resonantly enhancing
the gravitational wave signal sideband we performed numerical
simulations of the shot noise limited sensitivity using the
FINESSE\footnote{Our initial simulations did not include
radiation pressure nor optical spring effects. We later confirmed
with a separate analysis that these effects play no role for the
frequency band and in the configurations analyzed in this
paper.} software \cite{finesse}. As an example we have chosen an idealized and simplified
Michelson interferometer  with Power-Recycling employing tuned and
detuned arm cavities of 4\,km length. To simplify the simulation
of shot noise, a DC-readout scheme was used. Table \ref{LIGO_tab}
gives a summary of the simulation: With an input power of 4\,W a
circulating light power of 10\,kW is achieved in each arm cavity.
The corresponding shot noise limited displacement sensitivity of
such a configuration with arm cavities resonant for the carrier
light is shown in Figure \ref{Initial} (green solid trace).


\begin{table}[htbp]
    \begin{center}
        \begin{tabular}{|l|c|}
            \hline
            Transmission PRM & 10\,\%\\
            \hline
            Transmission ITX/ITY & 3\,\%\\
            \hline
            Transmission ETX/ETY & 0\,\%\\
            \hline
            Input light power at PRM & 4\,W\\
            \hline
            Light power in each arm & 10\,kW\\
            \hline
            Dark fringe offset at BS for DC-readout & 0.3\,deg  \\
            \hline
        \end{tabular}
        \caption{Simulation of an ideal  Michelson interferometer  with DC-readout.
        \label{LIGO_tab}}
    \end{center}
\end{table}
By shifting the microscopic position of ETX and ETY we can detune
the arm cavities, i.e. shifting their resonance. Here we choose a
detuning frequency of 200\,Hz corresponding to a detuning of
-1\,deg. This results in a reduction of the power buildup in the
arm cavities by a factor of 6.25 which can be compensated by
increasing either the input power from 4 to 25\,W or the
Power-Recycling gain (reducing the transmission of PRM from 10 to
1.7\,\%) in order to restore the nominal 10\,kW intra-cavity
power. Table \ref{para_tab} shows the parameters used for the
simulation of tuned and detuned arm cavity configurations.

\begin{table}[htbp]
    \begin{center}
        \begin{tabular}{|l|ccc|}
            \hline
            \hline
            Scheme   & Tuning   & Input & Transmission\\
                      &  ETX/ETY [deg] & power
                   [W] & PRM [\%] \\
            \hline
            \hline
            arm cavities resonant  & 90/0 & 4.0 &
            10.0\\
             for carrier & & & \\
            \hline
             detuned arm cavities,  & 89/-1 & 25.1 &
            10.0\\
            input power increased & & & \\
            \hline
             detuned arm cavities, & 89/-1 & 4.0 &
            1.7\\
             PR gain increased & & & \\
            \hline
            \hline
        \end{tabular}
        \caption{Parameters used for the simulation of tuned
        and detuned arm cavity configurations in an ideal Michelson interferometer. All three scenarios lead
        to an intra-cavity power of 10\,kW.
        \label{para_tab}}
    \end{center}
\end{table}

Figure \ref{Initial} shows the shot noise limited displacement
sensitivity of the two configurations with detuned arm cavities
(red dashed and blue solid line). In a frequency band of
between 150 to 350\,Hz an increased
sensitivity can be achieved. The maximal improvement is obtained
around 230\,Hz. Below 150\,Hz and above 350\,Hz the sensitivity of
the configurations with detuned arm cavities is worse than for
tuned arm cavities.

\section{Comparison of detuned arm cavities and resonant sideband
extraction} \label{sec:rse}

Resonant sideband extraction (RSE) \cite{rse} is a well known
concept for increasing the sensitivity of gravitational wave
detectors by placing an additional mirror in the output port of
the instrument (see Figure \ref{layouts} b). This concept is
foreseen to be implemented in the second generation instruments
such as Advanced LIGO \cite{advligo}. Similar to the concept of
using detuned arm cavities, RSE allows to increase the sensitivity
of the interferometer in a certain frequency band by sacrificing
the sensitivity outside this band.

Our simulations show that the sensitivity of a detector using
detuned arm cavities can exactly be reproduced by using tuned arm
cavities and RSE (see black solid line in Figure \ref{Initial}).
For the simulation of RSE we used a Signal-Recycling mirror (SRM)
with a reflectivity of 58\,\% and a tuning phase of 70\,deg. The
finding is that the following two concepts
\begin{itemize}
\item Tuned arm cavities together with RSE

\item Detuned arm cavities with increased Power-Recycling gain
\end{itemize}
are in principal equivalent.

Interestingly, the implementation of the second concept is
significantly easier. While RSE requires the installation of an
additional suspended mirror and a completely new control scheme
for additional degrees of freedom (longitudinal and alignment),
detuned arm cavities only require a slightly more complex length
sensing and control scheme (see below) and the exchange of PRM by one
with 1.7 instead of 10\,\% reflectivity\footnote{Both, Virgo and
GEO\,600, already replaced their PRM by one of higher reflectivity
during recent commissioning periods. Thereby no significant
problems had been encountered.}. Furthermore, the Fourier frequency
of the peak sensitivity can be adjusted online by tuning the
settings of the control system, similarly to an RSE system.

\begin{figure}[htb]
\centering
\includegraphics[width=0.7\textwidth]{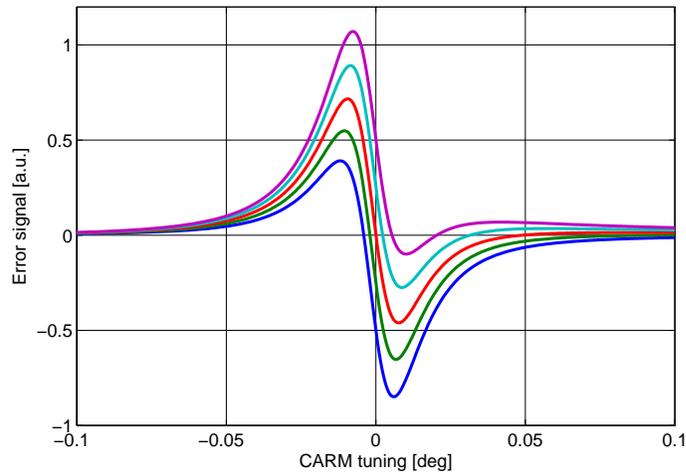}
\caption{Example error signal for the common mode arm cavity length (CARM) for
an interferometer with detuned arm cavities; zero CARM tuning refers to the new
detuned operating point.
The error signal is obtained by demodulating the
light reflected of the back of BS at an additional RF frequency.
The plot shows several traces for different demodulation phase (50 to 70
degrees, in 5 degree increments).} \label{errsig}
\end{figure}

 When implementing detuned arm cavities into a
currently running detector the bandwidth of the signal enhancement
(corresponding to the reflectivity of the SRM in RSE) is linked to
the Finesse of the arm cavities. A potential disadvantage of
detuned arm cavities might be the slightly higher light power
inside the small Michelson interferometer originating from the
increased Power-Recycling gain.

At first glance the length sensing and control for an
interferometer with detuned arm cavities looks very similar to the
standard LIGO control problem. Effectively only an offset to the
operating point of the common mode arm (CARM) signal has to be
introduced. Unfortunately this offset is much larger than the
dynamic range of the original CARM error signal. Hence, in order
to obtain a suitable CARM control signal a new RF modulation has
to be introduced. Figure~\ref{errsig} show example error signals
for the new CARM operating point derived from RF modulation at
approximately 15\,MHz. The demodulation phase in combination with
the exact value of the RF frequency determines the operating point
of the CARM signal. This poses a new constraint on the quality of
the control electronics, which however, is a known and understood
feature of detuned Signal-Recycling control systems as well
\cite{prototype}. Preliminary investigations show further that
appropriate error signals for all other degrees of freedom can
easily be obtained. The detuning of the arm cavities has virtually
no effect on the differential mode arm signal and the
Power-Recycling control but degrades slightly the quality of the
small Michelson error signal.

\section{Initial LIGO with detuned arm
cavities}\label{sec:real_LIGO}
\begin{figure}[Htb]
\centering
\includegraphics[width=0.8\textwidth]{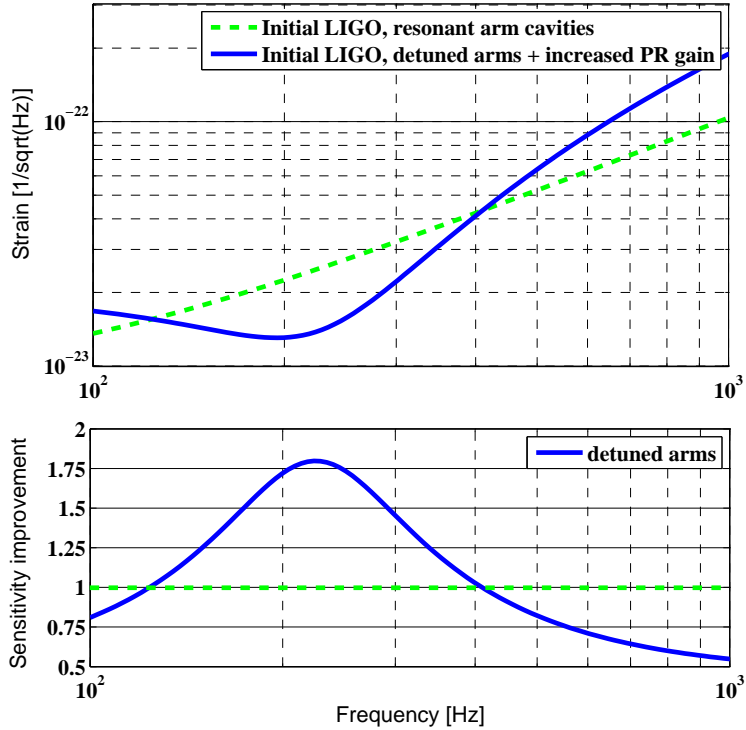}
\caption{UPPER SUBPLOT: Shot noise limited strain sensitivity for
initial LIGO using resonant arm cavities (green dashed trace) and
detuned arm cavities (blue solid line). In case of detuned arm
cavities the original PRM of 2.7\% transmittance was exchanged by
on with only 1\,\% transmittance. Still the circulating light
power is with 11\,kW smaller than for resonant arm cavities
(14.7\,kW). Using our proposed concepts allows to increase the
sensitivity of initial LIGO in the frequency band between 125 and
400\,Hz. LOWER SUBPLOT:  Ratio of the achieved strain
sensitivities with resonant and detuned arm cavities. A maximal
improvement of 80\,\% is achieved around 225\,Hz. } \label{ratio}
\end{figure}

 After we explained the principle
operation of the detuned arm cavity concept in the previous
section using an ideal Michelson interferometer with power
recycling, this section is devoted to simulating the benefit of
detuned arm cavities for a real system, i.e. initial LIGO.
  We used the best estimate for optical
parameter currently available \cite{rana}. The values for the most
important parameters are given in Table \ref{real_LIGO_tab}. The
sensitivity achieved with these parameters is shown in Figure
\ref{ratio} (green dashed line). Using resonant arm cavities in
the simulation yields an intra cavity power 14.7\,kW per arm.

This configuration is in the following compared to one optical
scheme employing detuned arm cavities. Since the sensitivity of
the initial LIGO detectors is only shot noise limited above
150\,Hz \cite{waldman}, we choose here a detuning frequency of
200\,Hz corresponding to a detuning of -1\,deg. In order to (at
least partly) compensate the decreased power buildup in the arm
cavities we use for this simulation a PRM with 1\,\% instead of
2.7\,\% transmission leading to an intra-cavity power of 11\,kW.
The corresponding shot noise limited sensitivity is shown in
Figure \ref{ratio} (blue solid line). In a frequency band between
125 and 400\,Hz an improvement is achieved. The maximal increase
in sensitivity amounts to about 80\,\% around 225\,Hz. Above
400\,Hz the decrease of the shot noise limited sensitivity
increases with the frequency. At 1\,kHz a sensitivity reduction of
45\,\% is observed.

\begin{table}[htbp]
    \begin{center}
        \begin{tabular}{|l|c|}
            \hline
            Transmission PRM & 2.7\,\%\\
            \hline
            Transmission ITX/ITY & 2.8\,\%\\
            \hline
            Transmission ETX/ETY & 10\,ppm\\
            \hline
            Loss at each coating & 90\,ppm\\
            \hline
            Radius of curvature ITX/ITY/PRM & -14.3\,km\\
            \hline
            Radius of curvature ETX/ETY & 8.0\,km\\
            \hline
            Input light power at PRM & 6\,W\\
            \hline
            Dark fringe offset at BS for DC-readout & 0.3\,deg  \\
            \hline
        \end{tabular}
        \caption{Simulation of initial LIGO with DC-readout. The parameters are taken from \cite{rana}.
        \label{real_LIGO_tab}}
    \end{center}
\end{table}

\section{Summary and outlook}
\label{sec:summary} We have introduced a new concept that uses
detuned arm cavities combined with a moderately increased
Power-Recycling gain to increase the shot noise limited
sensitivity of initial LIGO for frequencies between 125 and
400\,Hz, without increasing the light power inside the arm
cavities. This concept is found to give a sensitivity similar to
that of a conventional RSE configuration with a SRM of moderate
reflectivity. The implementation of detuned arm cavities might
require less new hardware than RSE. This is a considerable
advantage, however, further investigations and simulations should
be performed to evaluate the performance of the proposed scheme
with respect to noise couplings.

We have chosen to validate the concept using the initial LIGO
configuration because the real parameters and performance of the
running interferometers are well known. Nevertheless, the same concept
is applicable to these interferometers after the next upgrade: enhanced LIGO.
Further investigations are required, however, to understand the technical
limits to the circulating power in the arm cavities and the small Michelson
in a realistic enhanced LIGO scenario.

\ack{The authors would like to thank R.~Adhikari, M.~Evans, K.~Kawabe,
K.~Danzmann and H.~Lueck for fruitful discussions. A.~F. would like to thank
PPARC for financial support of this work. This document
has been assigned LIGO Laboratory document number
LIGO-P070067-00-Z.}


\section*{References}
\begin{thebibliography}{3}

\bibitem{geo}
Hild S \emph{et al} 2006 The Status of GEO\,600 \emph{ Class.
Quantum Grav.} \textbf{23}, 643--651.

\bibitem{virgo}
Acemese F \emph{et al} 2004 Status of VIRGO \emph{ Class. Quantum
Grav.} \textbf{ 21}, 385--394.

\bibitem{ligo}
Sigg D \emph{et al} 2004 Commissioning of LIGO detectors \emph{
Class. Quantum Grav.} \textbf{ 21} 409--415.

\bibitem{tama}
Takahashi R (the TAMA Collaboration) 2004 Status of TAMA300 \emph{
Class. Quantum Grav.} \textbf{ 21} 403--408.

\bibitem{malik}
Rakhmanov M, PhD-thesis, 2000, California Institute of Technology.

\bibitem{rehbein}
Rehbein H, personal communication 2007.

\bibitem{finesse}
Freise A  \emph{et al} 2004 Frequency-domain interferometer
simulation with higher-order spatial modes \emph{ Class. Quantum
Grav.} \textbf{ 21}, 1067--1074.

\bibitem{waldman}
Waldman S \emph{et al} 2006 Status of LIGO at the start of the
fifth science run \emph{Class. Quantum Grav.} \textbf{23},
653--660.

\bibitem{advligo}
www.ligo.caltech.edu/advLIGO/

\bibitem{rse}
Mizuno J \emph{et al} 1993  Resonant sideband extraction: A new
configuration for interferometric gravitational wave detectors
\emph{Phys. Lett. A} \textbf{175}, 273--276.

\bibitem{rana}
Adhikari R and Garofoli J personal communication 2007.

\bibitem{prototype}
Freise A \emph{et al} 2000 Demonstration of detuned dual recycling at the Garching 30 m laser interferometer
\emph{Phys. Lett. A} \textbf{277}, 135--142.

\end{thebibliography}
\end{document}